\documentclass[aps,pre,superscriptaddress,twocolumn,showpacs,nofootinbib]{revtex4-1}

\usepackage{amssymb}
\usepackage{amsmath}
\usepackage{amsthm}
\usepackage{graphicx}
\usepackage{amsfonts}
\usepackage{color}
\usepackage{natbib}
\usepackage{comment}
\usepackage{soul}
\usepackage{hyperref}
\usepackage[normalem]{ulem}

\hypersetup{
        colorlinks=true,
}
\newcommand{\ti}[1]{\ensuremath{_{\hspace{-0.5pt}\protect\raisebox{0pt}{\tiny{$#1$}}}}}

\newcommand{\bra}[1]{\langle #1|}
\newcommand{\ket}[1]{|#1\rangle}

\newcommand{\bea}{\begin{eqnarray}}
\newcommand{\eea}{\end{eqnarray}}
\newcommand{\myskip}[1]{}
\newcommand{\comments}[1]{}

\DeclareMathOperator{\tr}{tr}
\newcommand{\proj}[1]{}

\newcommand{\sectionprl}[1]{\par \textbf{#1.}---\!}
\newcommand{\subsectionprl}[1]{\par \textit{#1.}---\!}

\begin{document}
\title{Most energetic passive states}

\author{Mart\'i~Perarnau-Llobet}
\affiliation{ICFO-Institut de Ciencies Fotoniques, Mediterranean Technology Park, 08860 Castelldefels (Barcelona), Spain}
\author{Karen~V.~Hovhannisyan}
\affiliation{ICFO-Institut de Ciencies Fotoniques, Mediterranean Technology Park, 08860 Castelldefels (Barcelona), Spain}
\author{Marcus~Huber}
\affiliation{Departament de F\'{i}sica, Universitat Aut\`{o}noma de Barcelona, E-08193 Bellaterra, Spain}
\affiliation{ICFO-Institut de Ciencies Fotoniques, Mediterranean Technology Park, 08860 Castelldefels (Barcelona), Spain}
\author{Paul~Skrzypczyk}
\affiliation{$\mbox{H. H. Wills Physics Laboratory, University of Bristol, Tyndall Avenue, Bristol, BS8 1TL, United Kingdom}$}
\affiliation{ICFO-Institut de Ciencies Fotoniques, Mediterranean Technology Park, 08860 Castelldefels (Barcelona), Spain}
\author{Jordi Tura}
\affiliation{ICFO-Institut de Ciencies Fotoniques, Mediterranean Technology Park, 08860 Castelldefels (Barcelona), Spain}
\author{Antonio~Ac\'in}
\affiliation{ICFO-Institut de Ciencies Fotoniques, Mediterranean Technology Park, 08860 Castelldefels (Barcelona), Spain}
\affiliation{ICREA-Instituci\'o Catalana de Recerca i Estudis Avan\c cats, Lluis Companys 23, 08010 Barcelona, Spain}

\begin{abstract}
Passive states are defined as those states that do not allow for work extraction in a cyclic (unitary) process. Within the set of passive states, thermal states are the most stable ones: they maximize the entropy for a given energy, and similarly they minimize the energy for a given entropy. Here we find the passive states lying in the other extreme, i.e., those that maximize the energy for a given entropy, which we show also minimize the entropy when the energy is fixed. These extremal properties make these states useful to obtain fundamental bounds for the thermodynamics of finite-dimensional quantum systems, which we show in several scenarios.
\end{abstract}

\maketitle

\section{Introduction}

Passive states were introduced in \cite{pusz} as the ones obeying the second law of thermodynamics in the Kelvin-Planck formulation \cite{kpsl,ll5}. Namely, states that can yield no work in a Hamiltonian process at the end of which the system returns to its initial Hamiltonian, $H$. Any such process can be described by a unitary operation $U$, and if we define the maximal extractable work from the system as
\bea \label{WorkMax}
W_{\rm max}(\rho)=\max_U \tr \left[H\left(\rho - U \rho U^{\dagger} \right)\right],
\eea
then passive will be the states for which $W_{\rm max}=0$. The quantity $W_\text{max}$ was given the name ``ergotropy'' \cite{allah}.

Although the second law is formulated for thermal states \cite{gibbon}, passive states constitute a much wider class \cite{inverse}. In fact, they consist of all states that commute with the system Hamiltonian and have no population inversions \cite{pusz,lenard,allah}. Thermal states enter the picture in two ways. First, they are the ones that have the minimal energy for a given entropy. Second, thermal states are the only completely passive states. Complete passivity is another fundamental notion introduced in \cite{pusz}, and designates those states $\rho$ for which $\rho^{\otimes n}$ are passive for all $n$. The fact that \textit{only} thermal states are completely passive is very well illustrated by the elegant result in \cite{alicki}, stating that the asymptotically activatable work contained in a passive state $\sigma_p$, $W_\text{act}=\lim\limits_{n\to\infty}\frac{W_{max}(\sigma_p^{\otimes n})}{n}$, is given by
\bea \label{wact}
W_\text{act}=\tr(H\sigma_p)-\tr(H\tau_\beta),
\eea
where $\tau_\beta$ is the thermal state at the inverse temperature $\beta$ \cite{gibbon}, and $\beta$ is uniquely determined by requiring $S(\sigma_p)=S(\tau_{\beta})$. Here $S(\rho)=-\tr(\rho\ln\rho)$ is the quantum von Neumann entropy \cite{alicki}. Summarizing, provided it is not thermal, a passive state can be activated by jointly processing several copies of it, and the work that can be extracted in the limit of infinite copies is given by \eqref{wact}. 

Being motivated by the above results, the main goal of this work is to identify and study the other extreme of passive states (with respect to completely passive, thermal states), i.e., the ones that have maximal energy for a given entropy \cite{optimizations}. We refer to such states as the most energetic passive states (MEPS). We then show that, due to their extremal properties, these states provide useful information about fundamental thermodynamic processes. First, from their definition, it naturally follows that the MEPS have maximal activatable work content, see (\ref{wact}). Another motivation for our study is that while thermal states, when used instead of $U\rho U^\dagger$ in (\ref{WorkMax}), provide an upper bound on the extractable work, the MEPS provide a lower bound. From a methodological point of view, this gives a practical tool to estimate the usefulness of a given state from the perspective of average work extraction.

Akin to thermal states, the MEPS have a rather general characterization and are also monotonic with respect to entropy.
They constitute a one parameter family, and take a particularly simple form, 
\begin{equation}
\rho=\frac{\lambda}{k}\sum_{i=0}^{k}|e_i\rangle \langle e_i | + \frac{1-\lambda}{l}\sum_{i=0}^{l}|e_i\rangle \langle e_i |.
\end{equation}
where $|e_i \rangle$ are the energy eigenvectors, and $e_{i+1}\geq e_{i}$. That is, the state is (at most) a mixture of two projectors onto subspaces of states with energies lower than a given value. These states are known as $\theta$-canonical states \cite{berdi}, and are exactly the passive states related to microcanonical states. This gives a new meaning to this rarely used concept.


The MEPS  also allow us to quantify how energetically different passive and thermal states can be. Quite remarkably, it turns out that although the MEPS can deviate significantly from thermal states for different spectra, we give evidence that the MEPS of, e.g., many-body systems with short-range interactions, behave almost as thermal states and have little potential for locked (i.e., potentially activatable) work. This makes another case  for the universality of the thermodynamic formalism in the macroscopic world \cite{ll5,campa,brando,muma,kliesch}.

\section{Passive states}

Consider a process where the system remains thermally isolated: it can evolve according to its own Hamiltonian $H$ and due to external (time-dependent) fields, $V(t)$. Furthermore, the process is cyclic, i.e., the external fields are turned on and off at the beginning  and at the end, $V(0)=V(\tau)=0$. The corresponding evolution can be described by a unitary evolution $U$, with $U(\tau)= \overrightarrow\exp\left\{-i\int_0^\tau d t\left[H + V(t)\right]\right\}$.  Since the system remains thermally isolated, work is given by the change of its average energy, $W = \tr\left(\rho H \right)- \tr\left(U\rho U^\dagger H\right)$, where $H = \sum_{i} e_i \ket{e_i}\bra{e_i}$ ($e_{i+1}\geq e_i$) is the internal Hamiltonian of the system.


In this process, work is provided (or extracted) by the external time dependent fields $V(t)$. By appropriately choosing $V(t)$, we can generate every unitary operation $U$, and thus the operations considered in this context are equivalently all unitary operations. It follows that the maximal work that can be extracted from $\rho$ is given by \eqref{WorkMax}. This expression is maximized for $U \rho U^{\dagger}=\sigma_{p}$, with 
\begin{equation}
\sigma_p = \sum_{i} p_i \ket{e_i}\bra{e_i}, \quad
\text{with} \quad p_{i+1}\leq p_i,
\end{equation}
where $p_i$ are the eigenvalues of $\rho$ \cite{pusz,lenard}. In other words, given a state $\rho$, the (maximal) extractable work reads $W_{\rm max}=\tr (\rho H)-\tr (\sigma_p H)$ \cite{allah}.

\section{Main result}

In this section, for a given Hamiltonian $H$ and entropy $S$ \cite{entro}, we find the passive state that maximizes the energy, which we  denote by $\sigma_p^{\star}$. This maximization yields an upper bound on $W_{\rm act}$ in \eqref{wact},
\begin{equation}
W_{act} \leq \tr ((\sigma_p^{\star}-\tau_{\beta'}) H)\equiv \Delta_{\rm max}(S,E) 
\label{eqmax}
\end{equation} 
where $S(\rho)=S(\tau_{\beta'})=S$, and $E=\tr(H\rho)$. It is convenient to first consider the complementary optimization, i.e., to find the passive state that minimizes the entropy for a fixed energy $E$. We will show that both optimizations provide the same state.

It is useful to introduce the following set of $d$ linearly independent states:
\bea \label{legs}
\omega_k = \frac{1}{k} \sum_{i=1}^{k} \ket{e_i} \bra{e_i},\quad 1\leq k\leq d.
\eea 
Any passive state can be written as a convex combination of such states, $\sigma_p = \sum_{i=1}^{d} q_i \omega_i$, with $q_i\geq0$ and $\sum_i q_i=1$. Therefore, the set of passive states defines a convex polytope (in fact, a simplex) which we denote by $\mathcal{S}$, whose vertices are given by $\omega_k$ in \eqref{legs}.

Within $\mathcal{S}$, we are interested in the subset of states with constant energy, $\tr(\rho H)=E$. Since the energy $\tr(\rho H)$ is a linear function, the condition $\tr(\rho H)=E$ defines an hyperplane, which intersects with $\mathcal{S}$.  We denote by $\mathcal{SE}$ the polytope formed by this intersection, i.e., $\mathcal{SE}= \{\sigma_p\; : \; \sigma_p\in \mathcal{S},\tr(H\sigma) = E\}$.

\begin{figure}
   \includegraphics[width=\columnwidth]{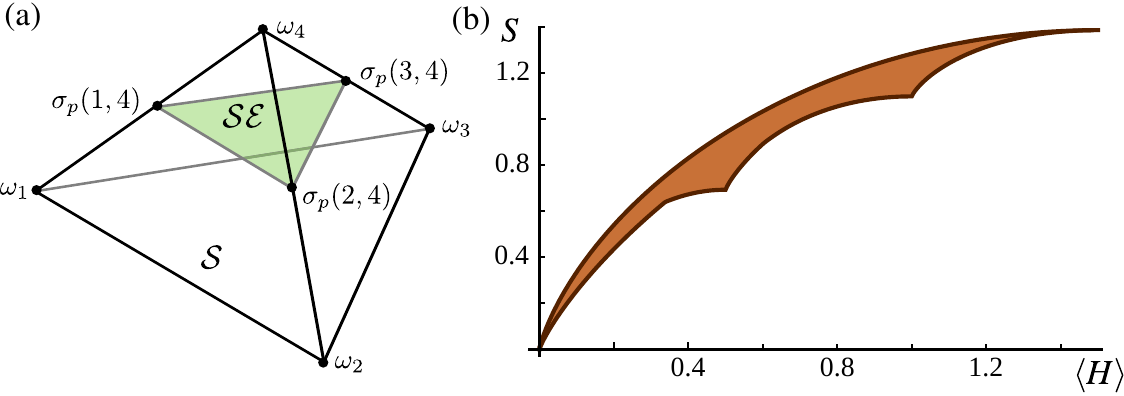}
    \caption{(a) The set of all passive states $\mathcal{S}$ and the intersection with the constant energy hyperplane $\mathcal{SE}$ for a four dimensional system. (b) Entropy versus energy given an equally spaced Hamiltonian of 4 levels, i.e., $H=\mathrm{ diag}\{0,1,2,3\}$. The shaded area corresponds to the simplex $\mathcal{S}$. The two boundary curves correspond to the set of thermal states (upper) and the most energetic passive states (lower).}
\label{f:polytope}
\end{figure}

The point then is to minimize the entropy function $S(\sigma)=-\tr(\sigma\ln\sigma)$ over $\mathcal{SE}$. These considerations are illustrated in Fig.~\ref{f:polytope}.
Since $\mathcal{SE}$ is a polytope and the entropy is a concave function, the minimum is achieved at a vertex \cite{rock}, each of which have a simple form \cite{Notedeg}. The vertices occur at the intersections of the energy hyperplane with the edges of $\mathcal{S}$ and therefore have the form $\sigma_p(k,l)=\lambda \omega_{k}+(1-\lambda)\omega_{l}$, where $\lambda=\lambda(k,l)$ is determined from the energy condition: 
\begin{equation}
\lambda(k,l)=\frac{\tr(H\omega_{l})-E}{\tr(H\omega_{l})-\tr(H\omega_{k})}.
\label{lambdakl}
\end{equation}
Note that for consistency $\tr(H\omega_{k})\leq E\leq\tr(H\omega_{l})$ must be satisfied, i.e., the vertices must be separated by the energy hyperplane. In general, the set of feasible index pairs $\mathcal{I(E)} = \{ (k,l) | \tr(H\omega_{k})\leq E\leq\tr(H\omega_{l})\}$ will depend on the spectrum of the Hamiltonian and the average energy $E$. It is however efficient to enumerate, with a system of dimension $d$ requiring only to check $O(d^2)$ pairs. The last step of the optimization is to minimize the entropy over all feasible pairs 
\begin{equation}
\label{sigmastar}
\sigma_p^{\star}(E)= \min_{\mathcal{I(E)}} \sigma_p(k,l),
\end{equation}
which can again be carried out efficiently for finite dimensional systems. 

We denote the entropy as $S^{\star}(E)\equiv S(\sigma_p^\star(E))$. If it is a monotonically increasing function of $E$, then $\sigma_p^{\star}(E)$ is also a solution of the complementary optimization, namely maximizing the energy when the entropy is fixed. In the following we show that this is the case by \textit{reductio ad absurdum}. We define the polytope of all passive states with an energy greater than or equal to $E$, $\mathcal{SE}_+(E)= \{\sigma_p\; : \; \sigma_p\in \mathcal{S}\;\text{and}\;\tr(H\sigma)\geq E\}$, whose vertices are those of $\mathcal{SE}$ plus those of $\mathcal{S}$ whose energies are at least  $E$. Again, the minimum of $S(\sigma)$ over $\mathcal{SE}_+(E)$, $S_+^\star(E)$, is achieved at one of the vertices. Assume that it is one of the $\omega_k$ with $\tr(H\omega_k)>E$. Consider the passive state $\alpha\omega_1+(1-\alpha)\omega_k$, with $\lambda$ given by $\lambda(k,1)$ in \eqref{lambdakl}, so that its energy is equal to $E$. A direct calculation shows that $S(\alpha\omega_1+(1-\alpha)\omega_k)<S(\omega_k)$, which contradicts our previous assumption. This implies that the minimum of $S(\sigma)$ over $\mathcal{SE}_+$ is attained on $\mathcal{SE}$, which, along with the observation that $\mathcal{SE}_+(E')\subset \mathcal{SE}_+(E)$ if $E'>E$, shows that the entropy is a nondecreasing function of $E$.

In conclusion, the passive states that maximize the energy for a fixed entropy, and at the same time minimize the entropy for a given energy, are the one parameter family defined by \eqref{sigmastar}. This family lies on the boundary of the set of passive states (see Fig.~\ref{f:polytope}), which are convex combinations of states given by \eqref{legs}. This suggests a beautiful relation within the set of passive states between canonical and $\theta$-canonical distributions: they give rise to the most and least stable states, respectively.

\section{Applications}

Besides the question of activation \cite{alicki,karen,nicolas}, there are other scenarios related to work extraction where the MEPS can be useful.

\textit{Weight}. The notion of passivity has been recently used to define the work in fully quantized heat engines \cite{davidI,davidII,davidIII}. In such set-ups, every component of the engine, including the weight which serves as a receiver of the extracted work, is a quantum mechanical system. Then it becomes natural to interpret any change in $W_\text{max}$, the work content of the weight, as the work exchanged with the engine \cite{davidI,davidII,davidIII}.

In order to relate the work stored in the battery with the standard notions of thermodynamics, such as the Carnot principle, it is useful to obtain bounds on \eqref{WorkMax} that do not depend on the whole spectrum of the state, but rather 
only on its energy and entropy. It is straightforward to see that \eqref{WorkMax} satisfies, $W_{\rm max} \leq \tr ((\rho-\tau_{\beta'}) H)$
where $S(\rho)=S(\tau_{\beta'})$. The authors of \cite{davidI,davidII,davidIII} use this bound to obtain an upper bound on the Carnot efficiency in fully quantized set-ups, where the entropy gain of the battery is non-negligible. Now, using the MEPS, we can find a bound on $W_{\max}$ in the other direction, i.e., $W_{\max} \geq \tr ((\rho-\sigma^{\star}_p) H)$, where $S(\rho)=S(\sigma_p^{\star})$. This expression provides a lower bound on the extractable work from a battery, which, again, only depends on the energy and the entropy of the state.


\textit{Thermal bath as an ancillary system}. Passive states can also be activated if one has an access to a thermal bath at some inverse temperature $\beta$. Then the second law of thermodynamics places an upper bound on the extractable work through the free energy difference of the system, $W\leq F_\beta[\sigma_p]-F_\beta[\tau^{(S)}_\beta]$, where $F_\beta[\rho]=\tr(H\rho)-\beta^{-1}S(\rho)$ is the free energy, and $\tau^{(S)}_\beta\propto e^{-\beta H}$ is the thermal state (see, e.g., \cite{ll5,paul,rodi,henrik,kmp}). The inequality can be saturated if the thermal bath is big enough and it is capable of thermalizing the system \cite{kmp}. 

Let us now define a fictitious thermal state, $\tau_{\beta'}^{(S)}$, whose temperature is adjusted to satisfy $S(\tau_{\beta'}^{(S)})=S(\sigma_p)$. We can now express the free energy difference as
\begin{equation}
F_\beta[\sigma_p]-F_\beta[\tau^{(S)}_\beta]=\Delta_{\max}(S,E_p)+F(\tau_{\beta'}^{(S)})- F(\tau_{\beta}^{(S)}).
\end{equation}
As in the previous case, this expression is maximized for $\sigma_p=\sigma_p^{\star}$, and thus $\sigma_p^{\star}$ provides a bound on the amount of work that can be extracted from a passive state using an external bath.

\section{Spectrum}

The amount of work $\Delta_{\max}(S,E_p^{\star})$ that can be locked in $\sigma_p^{\star}$ highly depends on the structure of $H$ and its dimension. As an extreme case, when the dimension $d$ of the system is 2,  all passive states are thermal and thus $\Delta_{\max}(S,E_p^{\star})=0$. As the dimension increases, so does $\Delta_{\max}(S,E_p^{\star})$, with a rate defined by the structure of $H$. In this section we give some general considerations in the limit of $d\rightarrow \infty$. These asymptotic results are then illustrated by  exactly solving some specific systems for finite dimensions.

\textit{Subexponential growth of the density of states with energy}. Let us assume a dense spectrum bounded from above by $E_m$ (the ground state is taken to be non-degenerate and to have zero energy). Assume that the density of states (DOS) \cite{DoS} scales polynomially with energy, $g_{\ti{E}}= c E^a$, where $c$ is some positive constant. The total number of states within $[0,E]$ is then given by $N_{\ti{E}}=\int_{0}^{E}\hspace{-1.5mm} dE' g_{\ti{E'}}=\frac{c}{a+1}E^{1+a}$. Let us define $\omega_{\ti{E}}$ as a state that is filled up to energy $E$, i.e., $\omega_{\ti{E}}\equiv \omega_{\ti{N_E}}$ in \eqref{legs}. It satisfies $\tr[\omega_{\ti{E}} H]\hspace{-0.5mm}=\hspace{-0.75mm}\frac{1}{N_{\ti{E}}}\hspace{-0.75mm}\int_0^E \hspace{-1.75mm} dE' g_{\ti{E'}}E'\hspace{-0.5mm}=\hspace{-0.5mm} \frac{a\hspace{-0.5mm}+\hspace{-0.5mm}1}{a\hspace{-0.5mm}+\hspace{-0.5mm}2}E$, and $ S(\omega_{\ti{E}})\hspace{-0.5mm}=\hspace{-0.5mm}\ln N_{\ti{E}}$. The MEPS is a combination of two such states,
$\lambda \omega_{\ti{E_1}}+(1-\lambda) \omega_{\ti{E_2}}$,
with $E_1$, $E_2$ depending on the specific case (entropy of the state, spectrum, etc). Numerical analysis provides evidence that $E_1=0$ and $E_2=E_\mathrm{m}$ is always the optimal choice for $N_{\ti{E}}\gg 1$. Therefore we focus on 
$\sigma_0=(1-\lambda) \ket{0}\bra{0}+\lambda \omega_{E_\mathrm{m}}$,
where $\lambda$ is determined by the energy (or entropy) of $\sigma_0$. 

The energy and entropy of $\sigma_0$ are given by $E(\sigma_0)=\tr[\sigma_0 H]= \lambda \frac{a+1}{a+2} E_\mathrm{m}$ and
$
S(\sigma_0)=H(\lambda)+\lambda \ln N_{\ti{E_\mathrm{m}}}+O\left(N_{\ti{E_\mathrm{m}}}^{-1}\right),$
where $H(\lambda)=-\lambda \ln \lambda - (1-\lambda) \ln (1-\lambda)$ is the binary entropy in natural units of information. From $E(\sigma_0)$ and $S(\sigma_0)$, one can express the entropy as a function of the energy, $S(E)$. Notice that 
$S(E) \rightarrow 0$ for $\frac{\ln E_m}{E_m}E \rightarrow 0$.
That is, if the norm of the Hamiltonian, $E_m$, is big enough, then essentially the state has zero entropy while having a finite energy. This is in sharp contrast with a thermal state, where if $S \rightarrow 0$ then $E\rightarrow 0$. This also implies that the energy hidden in a passive state $\Delta_{\max}(S,E_p^{\star})$ can be arbitrarily large in the $d\rightarrow \infty$ limit.  

\begin{figure}
   \includegraphics[width=0.9\columnwidth]{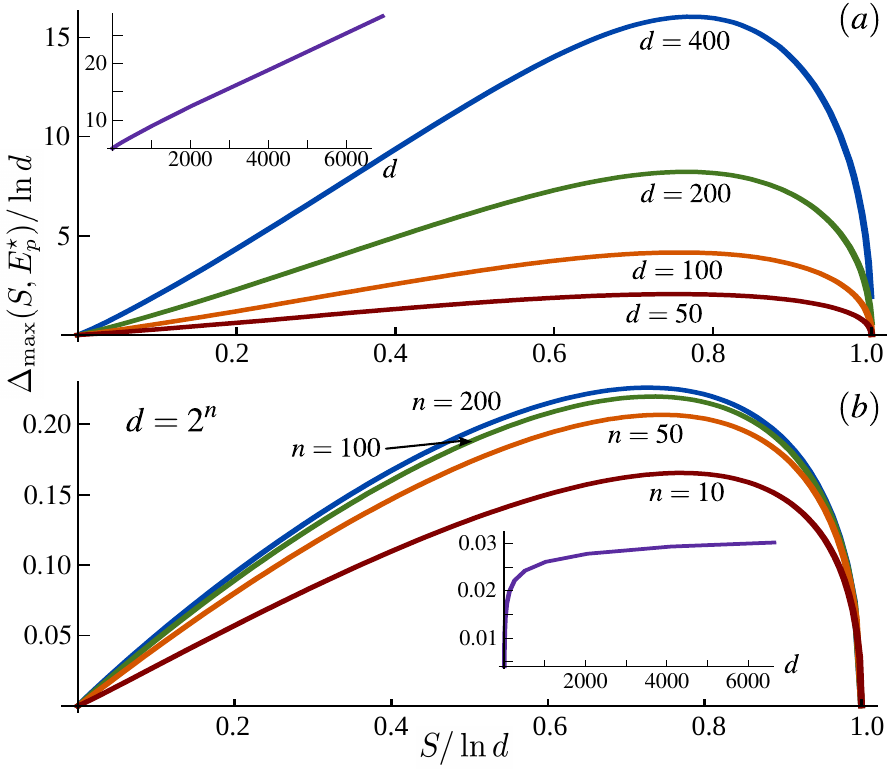}
    \caption{$\Delta_{\max}(S,E_p^{\star})/\ln d$ versus $S/\ln d$ (a) for an equally spaced Hamiltonian with $d=50,100,200,400$; (b) for a collection of $n$ non-interacting two level systems with $n=10,50,100,200$ (and $d = 2^n$). As the dimension increases, in (a) so does the energy difference between the most energetic passive state and the thermal state, while in (b) the difference grows much slower due to the presence of large degeneracies. Insets:  $\Delta_{\max}(S,E_p^{\star})/\ln d$ versus $d$ or $n$, for fixed small value of $S$. While in (a) there is linear growth, in (b) the value grows only logarithmically. }
\label{figDelta}
\end{figure}

\textit{Bathlike spectrum}. Assume that the DOS now scales as $g_{\ti E}=e^{bE}.$
Focusing again on $\sigma_0$, we obtain $E(\sigma_0)=\lambda \left(E_{m}-b^{-1}\right)+O\left(e^{-bE_m} \right)$.
From this expression we determine $\lambda(E)$ which, together with $N_E=(e^{bE_m}-1)/b$, can be inserted into $S(\sigma_0)$ to calculate $S(E)$. Taking, again, the limit $E/E_m \rightarrow 0$, we find $S \rightarrow bE$.
Here we observe that any amount of energy has an associated amount of entropy, even if the Hamiltonian is unbounded from above. Thus here, contrary to the previous case, the energy has a heatlike character.

To illustrate this behaviour, in Fig.~\ref{figDelta} we compute exactly $\Delta_\mathrm{max} (S,E_p^{\star})$ for a single system with equally spaced eigenvalues, $H=\sum_{k=1}^{d}k \ket{k} \bra{k}$, and for a collection of $n$ non-interacting two-level systems \cite{numberstates}. It is clearly observed how the presence of high degeneracies hinders the growth of $\Delta_\mathrm{max}(S,E_p^{\star})$ in the latter case, while in the former the energy difference grows with system size.

Finally, we note that an exponential scaling of the DOS with energy is a common assumption for the Hamiltonian of a bath. Indeed, a thermal bath is expected to be stable, has to thermalize any system that is brought in contact with it,  and has to be a large system with macroscopic (extensive) energy and entropy. On the other hand, the stability \cite{dunkel}, extensivity and the ability to thermalize \cite{ll5,kliesch,campa,brando,muma,arnau} hold only for systems with short-range interactions, which, in turn, have exponential DOS. 
Our results can thus be seen to provide a new insight on the role of the DOS. As opposed to polynomially growing DOS, passive states with exponential growth appear to behave pretty much like standard thermal states. This is in the spirit of the equivalence of canonical and microcanonical equilibria, that, again, holds only for systems with short range interactions \cite{ll5,campa,brando,muma}. It is worth adding that these type of spectra play an important role in fundamental questions such as  thermalization \cite{arnau} or the third law \cite{lluis}.

\section{Conclusions}

In this work we have characterized the family of passive states that maximize the energy of a system for a given entropy. We proved that they also solve the dual problem -- they minimize the energy for a given entropy. There is thus a clear parallelism with thermal states, which provide the reverse solution to such optimizations. These extremal properties make this class of states useful to obtain bounds in quantum thermodynamics of finite dimensional systems. Indeed, we have shown that this class provides a lower bound on the amount of work that can be extracted from a thermally isolated quantum system; and it places upper bounds on the extractable work from a set of passive states. 

We have also discussed how energy and entropy are related for the MEPS depending on the spectrum of the Hamiltonian. Whereas in thermal states any amount of energy is associated with some gain in entropy, we have shown that this is no longer true for \eqref{sigmastar} if the spectral density of the Hamiltonian increases sub-exponentially. This demonstrates a clean cut between bath-like spectra (collections of systems interacting with short ranged forces) and other types of spectra (systems with long range interactions).

Finally, the family of states found here can be used to lower bound the extractable work from a set of correlated  states, complementing the results in \cite{marti}. We leave as a future work to further explore the implications that the MEPS have for the efficiency of fully quantized heat engines \cite{davidI,davidII,davidIII}, for generalized notions of passive states \cite{funo,felo}, and for other scenarios where thermodynamic processes are modelled by unitary operations \cite{noah,spekkens,max,aberg,rodi,artur,reeb}.


\begin{acknowledgements}

This work is supported by the Spanish project FOQUS, the ERC CoG QITBOX, the EU project SIQS, the COST Action MP1209, and the Generalitat de Catalunya. M.P.L. also acknowledges funding from the Severo Ochoa program and the Spanish Grant No. FPU13/05988; M.H. from the Spanish MINECO through Project No. FIS2013-40627-P and the Juan de la Cierva fellowship (JCI 2012-14155), from the Generalitat de Catalunya CIRIT Project No. 2014 SGR 966, and from the EU STREP-Project “RAQUEL”; P.S. from the Marie Curie COFUND action through the ICFOnest program and ERC AdG NLST; and J.T. from the John Templeton Foundation, the Spanish Ministry project FOQUS (FIS2013-46768) and the Generalitat de Catalunya project 2014 SGR 874.

\end{acknowledgements}


\begin{thebibliography}{99}
\bibitem{kpsl} The second law of thermodynamics in Kelvin-Planck formulation: It is impossible to devise a cyclically operating device, the sole effect of which is to absorb energy in the form of heat from a single thermal reservoir and to deliver an equivalent amount of work. Here the device is general, i.e., it is the rest of the universe, and cyclicity is understood in therms of both the Hamiltonian and the state of the device \cite{ll5}.
\bibitem{gibbon} Thermal or, equivalently, Gibbs states are defined as $\tau_\beta=\frac{e^{-\beta H}}{\tr(e^{-\beta H})}$, where $H$ is the system Hamiltonian and $\beta>0$ is the inverse temperature.
\bibitem{inverse} In a way, the passive states are the solutions of the inverse problem of the second law (as in \cite{kpsl}). Indeed, while ``proving'' the second law, one typically derives the passivity for thermal states; here one requires the passivity and asks which are the states that have the property.
\bibitem{optimizations} Note that it is important that the optimisations are carried out restricted to the set of passive states. Otherwise they become trivial: the state with the least entropy for a fixed energy is a pure state, and the state with the most energy for a fixed entropy is a thermal state with a negative temperature.
\bibitem{entro} Let us stress that entropy is not an observable, and its value cannot be measured. What we assume here is that the knowledge of the entropy is somehow available. Even if one does not precisely know  the state, the entropy can be known. For example, think of a state preparator that mixes mutually orthogonal pure ensembles with given probabilities, but does not know which ensemble is at hand. In that case, only the spectrum of the resulting mixed state will be known, which is enough to calculate the entropy.
\bibitem{Notedeg} We ignore cases where the energy-hyperplane intersects the polytope on an entire face, since these cases are atypical and can always be removed by an $\epsilon$-perturbation of the the Hamiltonian or the energy.
\bibitem{DoS} The DOS is defined as the number of the eigenstates of the Hamiltonian per unit interval of energy.
\bibitem{numberstates} Note that the number of states increases as $N_E\propto e^{nH(p)}$ where $p$ is the local population of the states, while the energy is proportional to $n$.
\end{thebibliography}
\end{document}